\documentclass[conference]{IEEEtran}
\usepackage[T1]{fontenc}		% The two expressions are very important to load french accent
\usepackage[english]{babel}
\usepackage[latin1]{inputenc}	% The two expressions are very important to load french accent
\usepackage{listings}
\usepackage{color}

\usepackage{colortbl}
\usepackage{comment}
\usepackage{graphicx}
\usepackage{epsfig}
\usepackage{array, colortbl}
\newcolumntype{P}[1]{>{\raggedright\arraybackslash}p{#1}}
\usepackage{listings}
\usepackage{epstopdf}
\usepackage{multirow}
\usepackage{rotating}
\usepackage{subfig}
\usepackage{float}
\usepackage[obeyspaces,hyphens,spaces]{url}
\usepackage{balance}
\usepackage{fancybox}
\usepackage{scalefnt}
\usepackage[normalem]{ulem}
\usepackage{cleveref}

\usepackage[hang]{footmisc}
\usepackage{booktabs}
\usepackage{pbox}
\usepackage{ragged2e}
\usepackage{fancybox}
\usepackage{Libs/guehene}

\newenvironment{myindentpar}[1]%
{\begin{list}{}%
         {\setlength{\leftmargin}{#1}}%
         \item[]%
}
{\end{list}}

\newcommand{\hypobox}[1]{
	\begin{center}%
        \noindent\thicklines
        \setlength{\fboxsep}{6pt}%
        \setlength{\fboxrule}{0.7pt}%
        \fbox{
            \begin{minipage}{3.2in}%
            	\textit{#1}
            \end{minipage}
        }
	\end{center}
}

\makeatletter
\renewcommand{\paragraph}[1]{\noindent\textsf{#1}.}

\title{Stack Overflow: A Code Laundering Platform?}
\author{Le An, Ons Mlouki, Foutse Khomh, and Giuliano Antoniol
    \\
    SWAT--SOCCER Labs, Polytechnique Montr\'{e}al, Qu\'{e}bec, Canada
    \\
    \{le.an, ons.mlouki, foutse.khomh, giuliano.antoniol\}@polymtl.ca}

\begin{document}
\maketitle

\begin{abstract}
Developers use Question and Answer (Q\&A) websites to exchange knowledge and expertise. Stack Overflow is a popular Q\&A website where developers discuss coding problems and share code examples. Although all Stack Overflow posts are free to access, code examples on Stack Overflow are governed by the \emph{Creative Commons Attribute-ShareAlike 3.0 Unported} license that developers should obey when reusing code from Stack Overflow or posting code to Stack Overflow. In this paper, we conduct a case study with 399 Android apps, to investigate whether developers respect license terms when reusing code from Stack Overflow posts (and the other way around). We found 232 code snippets in 62 Android apps from our dataset that were potentially reused from Stack Overflow, and 1,226 Stack Overflow posts containing code examples that are clones of code released in 68 Android apps, suggesting that developers may have copied the code of these apps to answer Stack Overflow questions. We investigated the licenses of these pieces of code and observed 1,279 cases of potential license violations (related to code posting to Stack overflow or code reuse from Stack overflow). This paper aims to raise the awareness of the software engineering community about potential unethical code reuse activities taking place on Q\&A websites like Stack Overflow. % play in  potential misconduct that are taking place developers   %Overall, we observe However, we do not observe that developers respect the copyright terms in these code reusing candidates, which can lead to license violations. The results of this paper aims at reminding developers to pay attention to the copyright terms when reusing code from the Internet.
\end{abstract}


\IEEEpeerreviewmaketitle

\section{Introduction}
\label{sec:intro}
Question and Answer (Q\&A) websites, such as \emph{Stack Overflow}\footnote{\url{http://stackoverflow.com}}, allow users to share knowledge and expertise. These websites have become large knowledge repositories for developers to communicate on technical problems and resolve programming issues.
%When a developer encounters an issue, she might go to a Q\&A website to ask a question or seek relevant discussions. Once she finds a valuable answer, she might reuse a piece of code from the website in her own project.
%\ANLE{I removed some sentences here}
However when reusing code from Stack Overflow, developers should comply with the license of the code. Software licenses govern the use or redistribution of software. A failure to comply with a license term can result in bitter legal battles and large fines, as evidenced by the legal battle between Google and Oracle over nine lines of code~\cite{GoogleOracle}. Stack Overflow applies the \emph{Creative Commons Attribute-ShareAlike} (CC BY-SA 3.0) license~\cite{cc-by-sa} to restrict the usage of its content. Therefore, developers who violate the terms of this license when reusing code from Stack Overflow are exposed to penalties. Also, if a developer copies code from an existing software system and shares it on a Q\&A website (like Stack Overflow) without citing the reference, she would also be violating the software's copyright terms. Copying code from Stack Overflow to a software system or the other way around can lead to license violations and developers could be sued by the code owners.
%
%Software licenses govern the use or redistribution of software. A failure to comply to a license term can result in bitter legal battles and large fines, as evidenced by the legal battle between Google and Oracle over 9 lines of code~\cite{GoogleOracle}. Besides open-source software systems, developers should also respect the copyright terms to reuse content from Q\&A websites. For example, Stack Overflow applies the \emph{Creative Commons Attribute-ShareAlike} (CC BY-SA 3.0) license~\cite{cc-by-sa} to restrict the usage of its content. However, in real life, developers may not pay attention to the restrictions and directly copy code from Stack Overflow to a software system or the other way around. Both operations can lead to license violations and developers could be sued by the code owners.

In a previous work, Sojer et al.~\cite{sojer2011license} observed that developers do not always check copyright terms thoroughly when reusing code from Internet accessible open-source software. They also observed that some developers intentionally ignore the obligations imposed by licenses when reusing code from open-source software~\cite{sojer2011license}. In a recent study~\cite{mlouki2016detection}, we found 17 Android apps with license violations; suggesting that the developers of these apps disregarded the legal constraints of licenses' terms when reusing code from third-party sources in their software. However, both of these studies did not investigate the role that Q\&A websites could have played in these license violations. Yet, copy-paste operations from (and to) Stack Overflow can also lead to license violations. In particular, although Stack Overflow is free to access and its content can be easily searched by Google, developers seem to have less knowledge about the restrictions of Stack Overflow, in comparison to other software systems; as illustrated by this discussion on Stack Exchange~\cite{license_discussion}.
%\ANLE{I removed some sentences here}
%\red{In other words, some developers might directly copy code from Stack Overflow to their software without using the CC BY-SA 3.0 license, or copy code from an open-source system to a Stack Overflow post without citing the reference and--or providing the original license.}

In this paper, we conduct a quantitative study to investigate whether developers respect license restrictions when reusing code from Stack Overflow to Android apps, or posting the code of an Android app in a Stack Overflow question. We analyze 79.2k files extracted from 399 apps and 2.1M Stack Overflow posts that are related to Java and Android questions. %extracted from July 2008 to March 2016.
We use a state-of-the-art clone detection tool~\cite{svajlenko2014evaluating} NiCad~\cite{cordy2011nicad}, to identify duplicate code between the two studied datasets (\ie{} Apps' code and Stack Overflow posts). To ensure that code clones reported by NiCad are real code clones, we manually validate all occurrences of code clones found between the two datasets. We answer the following four research questions:

\newcommand{\RQone}{Do developers release apps with code copied from Stack Overflow?}
\newcommand{\RQtwo}{Do developers respect the copyright terms of code reused from Stack Overflow?}
\newcommand{\RQthree}{Do Stack Overflow users respect copyright terms when publishing code snippets on Stack Overflow?}
\newcommand{\RQfour}{How long does a Stack Overflow code snippet remain in released versions of an app?}

\begin{description}
\item[\textit{RQ1:}] \textit{\RQone}
\end{description}
\begin{myindentpar}{0.5cm}
In the 399 subject apps, we found 232 Android code snippets that are exact clones of code snippets posted on Stack Overflow. These code snippets are distributed in 135 files from 62 different apps. This result provides a quantitative evidence of potential code copying from Stack Overflow to Android apps.
%Most of these code snippets are ``reused'' from the question post in a discussion thread, and are related to the design of the Android user interface.
\end{myindentpar}

\begin{description}
\item[\textit{RQ2:}] \textit{\RQtwo}
\end{description}
\begin{myindentpar}{0.5cm}
We investigated the licenses of the 232 code snippets that were potentially reused from Stack Overflow, %with potential that reused code Among the code In the code reuse candidates detected in \textbf{RQ1},
and observed potential cases of license violations in 60 apps. %\Foutse{we should elaborate more}.% containing code reuse candidates carry a risk of software license violations and legal punishment.
We contacted the developers of the apps in which the violations were found and received some confirmation of code reuse from Stack Overflow, with one developer saying: \emph{``there is definitely code in our project that is copy-pasted from Stack Overflow, as I have done this several times. I assumed (falsely it seems) that everything there is public domain''}. %Another developer reported that the copied code \emph{``was inherited from (another project)''}. These replies suggest that
These results are an indication of potential unethical code reuse from Stack Overflow. %According to the replies, some developers do not handle the issue of license violations seriously.
Software organizations should consider putting in place license control and management mechanisms to avoid exposing themselves to license violation issues.
%In fact, one of our survey developers argued for it: \emph{``We don't have any policy about that. Now might be a good time to have that discussion."} %Developers seem These  Since asking and answering questions on Q\&A websites is nowadays a common practice, software organization should consider putting in place mechanisms that ensures ethical code reuse in their software. We also advocate for the development of tools to assist developers in license  managements.

%raise developers' awareness about the copyright terms of Q\&A websites and remind them to correctly use software licenses.
\end{myindentpar}

\begin{description}
\item[\textit{RQ3:}] \textit{\RQthree}
\end{description}
\begin{myindentpar}{0.5cm}
We observe 1,226 Stack Overflow posts potentially reusing code from respectively 68 Android apps. 1,219 of these posts have a potential risk of license violations. A majority (83.9\%) of the large code snippets (with more than 50 lines) contained in these Stack Overflow posts are related to the \emph{Android Navigation Drawer} component.  
We also found 126 code snippets that seem to have migrated from one app to Stack Overflow and then from Stack Overflow to another app. In 12 of the migrated code snippets, the file containing the code snippet in the first app and the file containing the code snippet in the second app use different software licenses.
%In total, these migrated code snippets appeared in \Foutse{are you sure that 126 code snippets appeared in 1 219 posts? you didn't mentioned this in the result section of rq3!!!}1,219  %, suggesting potential license violations.} % are likely to violate the corresponding apps' licenses.
\end{myindentpar}

\begin{description}
\item[\textit{RQ4:}] \textit{\RQfour}
\end{description}
\begin{myindentpar}{0.5cm}
Most of the code reused from Stack Overflow remained in the apps for up to 20 releases. In some cases, the code remained in the app for more than 300 releases and--or during a period of more than four years. The fact that these code snippets with potential license violations remained in the apps for such a long time suggests that some developers do not pay enough attention to copyright terms. %of Stack Overflow. %Developers should be careful when reusing code from Stack Overflow and software organizations should check the license terms of their code prior to their release. % assurance teams should to check the licenses of the reused code to avoid or remove license violations as soon as possible.
\end{myindentpar}

Overall, this paper makes the following contributions:
\begin{itemize}
\item To the best of our knowledge, this is the first quantitative study about the misuse of software licenses on a large Q\&A website.
\item We provide quantitative and qualitative evidences of potential unethical code reuse on Stack Overflow. We hope that the results of this study will raise the awareness of the software community about license issues in Q\&A websites, which are now very popular in developers' communities.
%\item We leverage two large-scale datasets, 2.1M Stack Overflow posts and 79.2k Android files, to conduct our case study.
%\item We do not observe that developers provided any link to the original code or applied the appropriate license for the code reuse candidates.
%\item \emph{Android Navigation Drawer} is very intensively discussed when developer share large code chunks to Stack Overflow.
\end{itemize}

\textbf{The remainder of this paper is organized as follows.}
Section \ref{sec:back} discusses license restrictions in open-source software and Stack Overflow. Section \ref{sec:design} describes the design of our case study. Section \ref{sec:case-study} presents the results of the case study. Section~\ref{sec:discussion} discusses threats to the validity and the contributions of this study. Section~\ref{sec:related} summarizes related works and Section~\ref{sec:conclusion} concludes the paper.

\section{License Restrictions in Open-Source Software and Stack Overflow}\label{sec:back}

In this section, we discuss general license restrictions in open-source software and Stack Overflow.
Open-source software licenses allow free access to the source code of a software system. However, reuse and--or distribution
are often limited by certain restrictions~\cite{open_licenses}. Most open-source software licenses possess different versions, each of them with their own restrictions. There exist two kinds of open-source licenses: \emph{restrictive} licenses (also known as ``copyleft'' or ``reciprocal'' licenses) and \emph{permissive} licenses~\cite{vendome2015and}. Restrictive licenses enforce restrictions on the license of derivative works. For example, the GPLv3.0 license says this about derivative works: \emph{``You must license the entire work, as a whole, under this License to anyone who comes into possession of a copy".} %require developers to use the same license to distribute a new software that integrates software licensed under a restrictive license (\eg{} \emph{GPL} license). 
However, permissive licenses allow software distribution under a different license (\eg{} BSD and MIT licenses). %\red{In general, reusing code into a new software system with incompatible licenses can lead to license violations.}
%For example, the GNU General Public License (GPL) restricts the modification and redistribution of the code. It is stated that one : ``must license the entire work, as a whole, under this License to anyone who comes into possession of a copy''~\cite{GPLlicense}. The Apache Software Foundation also requires that all \emph{Apache v2} software must be distributed under the \emph{Apache v2} license or the \emph{GPL v3} license.
%Also, for GPL license, the Section 5 of the \emph{GPL v3.0} license~\cite{GPLlicense} restricts the code modification and redistribution under this license: ``You must license the entire work, as a whole, under this License to anyone who comes into possession of a copy''.
%On the other hand, permissive licenses allow software distribution under a different license, \eg{} BSD and MIT licenses~\cite{singh2012networks} are compatible. In general, the reuse of different pre-existent components into a new software system can lead to license violations.
%In \Foutse{please rephrase the following paragraph...it is hard to understand....fix the english!}March 2015, VMware was sued over an improper use of the Linux kernel~\cite{VMware}, which is under the \emph{GPL v2} license, into its proprietary product. However the GPL licensed code should not be reused into proprietary software, because it requires the software releasing code under the same license.

In Stack Overflow, all user-generated content is licensed under the \emph{Creative Commons Attribute-ShareAlike 3.0 Unported} license (CC BY-SA 3.0)~\cite{cc-by-sa}. Under this license, users can share and adapt the content in the website, but they must respect the following restrictions~\cite{cc-by-sa_summary}:
\begin{itemize}
\item Attribution: \emph{``You must give appropriate credit, provide a link to the license, and indicate if changes were made. You may do so in any reasonable manner, but not in any way that suggests that the licensor endorses you or your use.''}
\item ShareAlike: \emph{``If you remix, transform, or build upon the material, you must distribute your contributions under the same license as the original.''}
\end{itemize}

In general, if a user copies a code snippet from Stack Overflow and reuses it into her own project, she must provide a reference to the original material and a link to CC BY-SA 3.0. She should also indicate any changes in case of a derivative. In addition, she can only share or release these projects under the CC BY-SA 3.0 or its later versions.

\section{Case Study Design}\label{sec:design}

In this section, we describe the data collection and analysis approaches that we use to answer our four research questions.

%\begin{enumerate}
%\item \RQone
%\item \RQtwo
%\item \RQthree
%\item \RQfour
%\end{enumerate}

\begin{figure*}[t]
\centering
\includegraphics[width = 15cm]{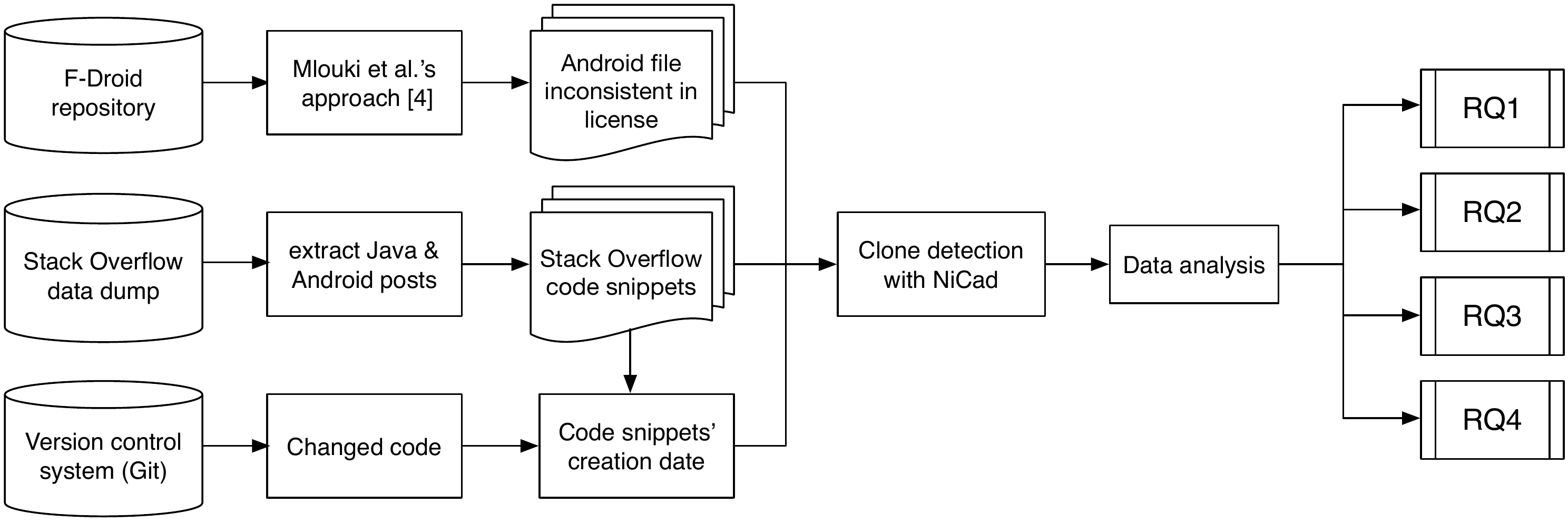}
\caption{Overview of our data processing approach.}
\label{fig:flow_graph}
\vspace{-10pt}
\end{figure*}

Figure \ref{fig:flow_graph} shows a general overview of our data processing approach. We describe each step in our data processing approach below. The corresponding data and scripts are available online at: \url{https://github.com/swatlab/stack_overflow}.

\subsection{Data Collection}\label{data_collection}
In our previous work~\cite{mlouki2016detection}, we found 399 apps with license inconsistences (\ie{} files that share similar code but having different licenses). % We leverage Mlouki et al.'s ``inconsistent files'' ~\cite{mlouki2016detection}, which are extracted from 857 Android applications in the F-Droid repository. Distributed in 399 applications,
%the F-Droid repository from ??? to ???\ANLE{the studied period}.
%these inconsistent files can be grouped into license inconsistent pairs. \ANLE{Please check if we can express like this, or just say the files are extracted from 399 apps?}
In this paper, we leverage this dataset of 399 apps to investigate the role that the Q\&A website, Stack Overflow, could have played in the occurrence of these license inconsistencies. We focus on files with license inconsistencies because they are likely to cause license violations. In total, the 399 apps contain 79,222 files with inconsistent licenses, which account for 1.4GB.
%
%In each pair, the two files contain a similar code snippet but use different software licenses. We focus on these files, because they are likely to introduce software license issues. In total, there are 79,222 inconsistent files, which account for 1.4GB.
We intend to investigate whether these files contain any code snippet reused from--or--to Stack Overflow. Stack Overflow shares its data in XML format as part of the Stack Exchange data dump\footnote{https://archive.org/details/stackexchange}, which is updated every three months under the CC BY-SA 3.0 license~\cite{cc-by-sa}. In this paper, we study Stack Overflow's data dump from July 2008 until March 2016.

\subsection{Preprocessing of Stack Overflow Data}
Stack Overflow's posts are stored in the \texttt{Posts.xml} file of the Stack Exchange data dump. \texttt{Posts.xml} accounts for 42GB.
%If we use an XML parser to analyze this file, it requires tremendous computing resources. Instead,
We write a Python script to identify posts from this file. From each post, we use regular expressions to extract the following information:
\begin{itemize}
\item \emph{Post ID}: the identifier of a post. Different posts in the same discussion thread possess different IDs.
\item \emph{Post creation date}: the submission date of a post.
\item \emph{Post tags}: the topics of a discussion thread, such as programming languages (\eg{} Java, C++), development environments (\eg{} Eclipse), and deployment platforms (\eg{} Android, iOS).
\item \emph{Post body}: the content of a post, which keeps the HTML format as in Stack Overflow's website.
\end{itemize}

In this paper, we only study posts with \texttt{Java} and \texttt{Android} tags, because we will compare the similarity of code snippets extracted from Stack Overflow with the source code of Android apps, and the majority of code in Android apps is written in Java. We use the creation date of Stack Overflow posts to decide whether a code snippet appeared on Stack Overflow before or after the creation of its corresponding clone in an Android app. %from is submitted earlier than its duplicate Android code.
If a code snippet was posted on Stack Overflow before the apparition of its clone in an Android app, we consider it as a reuse candidate from Stack Overflow to the Android app, meaning that the developers of the Android app probably reused the code from a Stack Overflow post. If the code snippet appeared in the App first before it was posted on Stack Overflow, we consider it to be a code reuse candidate from the Android app to Stack Overflow. Since the body of a post is kept in HTML format, we use the following regular expressions to extract source code snippets from the post:\\

\texttt{<pre><code>(.+?)</code></pre>}\\

\noindent
We save each extracted code snippet into a separate file. We eliminate the snippets with less than 10 lines of code, because too few lines of code can lead to noises in clone detections. In total, 2,106,303 code snippets are considered in our study, which account for 8.6GB.

\subsection{Clone Detection}\label{clone}
We use the clone detection tool, NiCad~\cite{cordy2011nicad}, to identify duplicate code between the studied source code datasets, \ie{} Android app dataset and Stack Overflow dataset. NiCad can detect \emph{Type 1} (exactly similar code snippets), \emph{Type 2} (syntactically similar code snippets), and \emph{Type 3} (copied code with further modifications) clones~\cite{roy2009comparison}. It can handle source code written in multiple languages, such as Java, C++, and Python. Svajlenko et al.~\cite{svajlenko2014evaluating} compared the performance of 11 clone detection tools from the literature and reported that NiCad achieved higher precision and recall, in comparison to the other 10 clone detection tools. %is assessed to be able to achieve a high clone detection accuracy by .
In addition, NiCad's cross-project clone detection feature allows us to only detect code clones between the two datasets instead of within each dataset. Since both studied datasets are very large in size and clone detection is a very resource consuming process, the cross-project clone detection feature of NiCad is useful to reduce the cost of the clone detection process, allowing us to analyze large code repositories. %us reduce the overall clone detection complexity than input the whole datasets into NiCad; providing us the feasibility to analyze large-scale code repositories.
We use the default settings of NiCad, \ie{} each clone pair has more than 70\% of similarity and the clones contain at least 10 lines of code.

During the clone detection process, NiCad requires an analytic memory space whose size is often 50 times larger than the analyzed dataset. Considering the size of our two datasets (\ie{} Android apps (79.2k files, 1.4GB) and Stack Overflow code snippets (2.1M files, 8.6GB)), we cannot feed the whole dataset into NiCad. Consequently, we split both datasets into slices. We limit the size of each Stack Overflow slice to 2,000 code snippets. Thus, the Stack Overflow data are split into 55 subsets. Each subset accounting for 160MB in average. Similarly, we split the Android dataset into 100 subsets, where each subset accounts for 14MB in average. We deliberately set each Stack Overflow slice larger than each Android slice, because firstly, NiCad will automatically filter out some irrelevant code such as %code with less than 10 lines or
code not corresponding to the syntax of Java. In addition, we tuned the split number for both studied datasets, the current splitting strategy allows NiCad to provide results faster.

Next, we perform clone detection with NiCad for $55 \times 100 = 5,500 $ rounds. We write a Python script to automate these clone detection rounds, \ie{} when one round is finished, the next round will be automatically started. Finally, we combine the results of each of the subsets as the total results of the clone detection between the two studied datasets. In this paper, we leveraged multiple computers with 32GB or 64GB memory, and finished the whole clone detection process in more than one month (including reprocessing for some failed clone detection rounds).

\section{Case Study Results}\label{sec:case-study}
This section presents and discusses the results of our four research questions. For each question, we describe the motivation, the approach followed to address the question, and the findings. To simplify the text, we define the following terms:

\begin{itemize}
\item App: one of our studied Android applications.
\item Post: One of our studied Stack Overflow post. We refer to the author of a post as a poster.
\item Similar: Two pieces of code are \emph{similar} if they are identified as being clones by NiCad (with its default parameters). %identified them as being clones.
%\item Code reuse candidate: if two code snippets are \emph{similar}, we call the later created snippet as a code reuse candidate.
\end{itemize}

\subsection*{RQ1: \RQone}
\noindent
\emph{\textbf{Motivation.}}
%Based on our experience and intuition, developers often seek references and answers in Stack Overflow when they encounter programming issues. If a piece of code in Stack Overflow can address a developer's issue, the developer may copy the code to her own project (sometimes with little modification).
Stack Overflow allows developers to ask and answer questions about programming problems~\cite{barua2014developers}. If a piece of code from Stack Overflow addresses a developer's issue, she may reuse the code in her project (sometimes with little modification). In this preliminary question, we look for evidences of code reuse from Stack Overflow in Android apps. % to our subject Android apps.
%investigate whether developers copy Java code from Stack Overflow to our studied Android files.
We are interested in understanding the role that Stack Overflow could have played in the occurrence of the license inconsistencies observed in our previous study~\cite{mlouki2016detection}. %This preliminary question can help us investigate the subsequent problems, such as whether developers respect software licenses when they reuse code from--to Stack Overflow.
%In addition, since we use Mlouki et al.'s ``license inconsistent files'' as our studied Android files~\cite{mlouki2016detection}, we want to explore whether code copied from Stack Overflow can lead to license inconsistency. Because when different developers copy code from the same Stack Overflow post that does not provide a license, the developers may apply different licenses to the code, which can lead to license incompatibility.

\noindent
\emph{\textbf{Approach.}}
As described in Section~\ref{clone}, we use NiCad to identify code clones between Stack Overflow posts and Android app files. For a given clone pair, if the Android code was created later than the Stack Overflow code, we consider that the cloned code was reused from Stack Overflow to the Android app and flag the Android code as a ``code reuse candidate''. To identify the creation date of a clone snippet from an app's file, we compare the clone snippet against the whole revision history of the file. We write a Python script to automatically match a cloned snippet to each \emph{added} line in the corresponding file's changing commits in Git. We note the date of the earliest matched commit as the creation date of the code snippet from the Android app. In total, we found 434 Android code snippets that are \emph{similar} to a Stack Overflow code snippet. 346 of these Android code snippets' creation date can be automatically identified. For the remaining 88 Android code snippets, we manually reviewed their files' revision history to identify the creation dates of the cloned code snippets. %r cloned code's creation date.

To determine whether a code snippet reused from Stack Overflow can lead to a license inconsistency, we proceed as follows. First, we find the snippet's corresponding file and locate the snippet's line numbers ($Line_{clone}$) in the file. Next we identify the line numbers ($Line_{inconsist}$) of the portion of code in this file, which is concerned by the license inconsistency (following the same approach as in our previous work~\cite{mlouki2016detection}). %, we then seek the file's inconsistent line numbers ($Line_{inconsist}$) and
Then, we calculate the common line numbers ($Line_{common}$) in $Line_{clone}$ and $Line_{inconsist}$. Finally, we compute the rate of $Line_{common}$ over $Line_{inconsist}$. We refer to this rate as \emph{overlapped rate} in the rest of the paper. If the \emph{overlapped rate} is greater than 0, it means that the reused candidate (or a part of the reuse candidate) is concerned by the file's license inconsistency issue; implying that this reused code can lead to a software license violation issue.

\begin{figure}[t]
\centering
\includegraphics[width = \columnwidth]{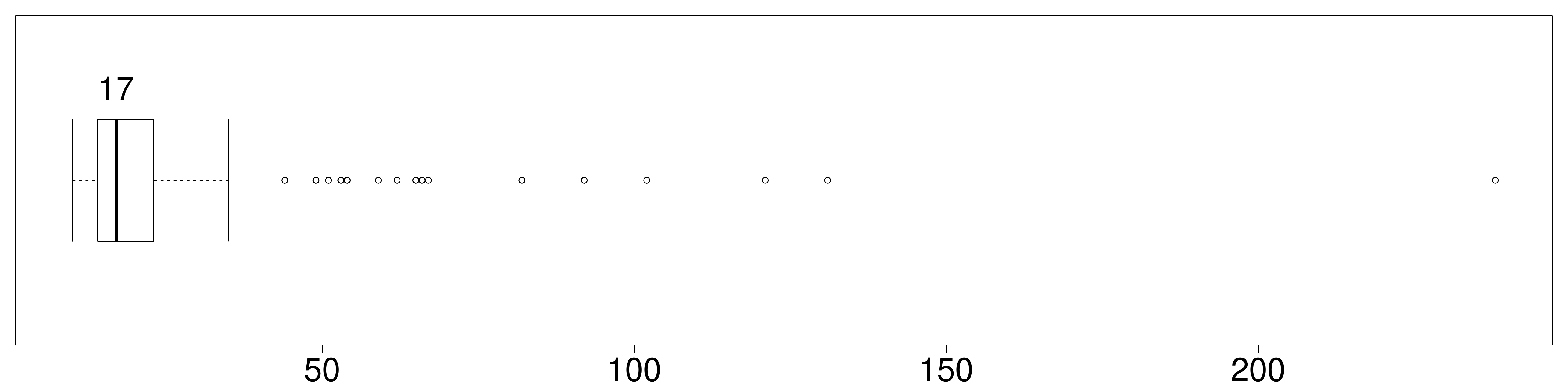}
\caption{Number of lines of an Android code snippet similar to a Stack Overflow post.}
\label{fig:app_clone_lines}
\vspace{-10pt}
\end{figure}

\begin{figure}[!]
\centering
\includegraphics[width = \columnwidth]{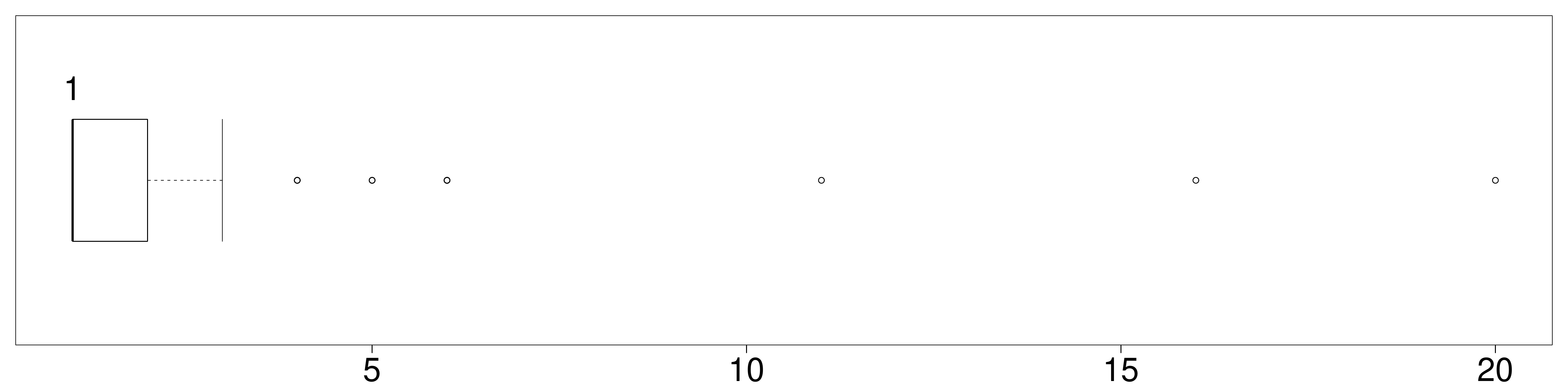}
\caption{Number of Android code snippets similar to the same Stack Overflow post.}
\label{fig:apps_from_a_post}
\vspace{-10pt}
\end{figure}

\begin{figure}[!]
\centering
\includegraphics[width = \columnwidth]{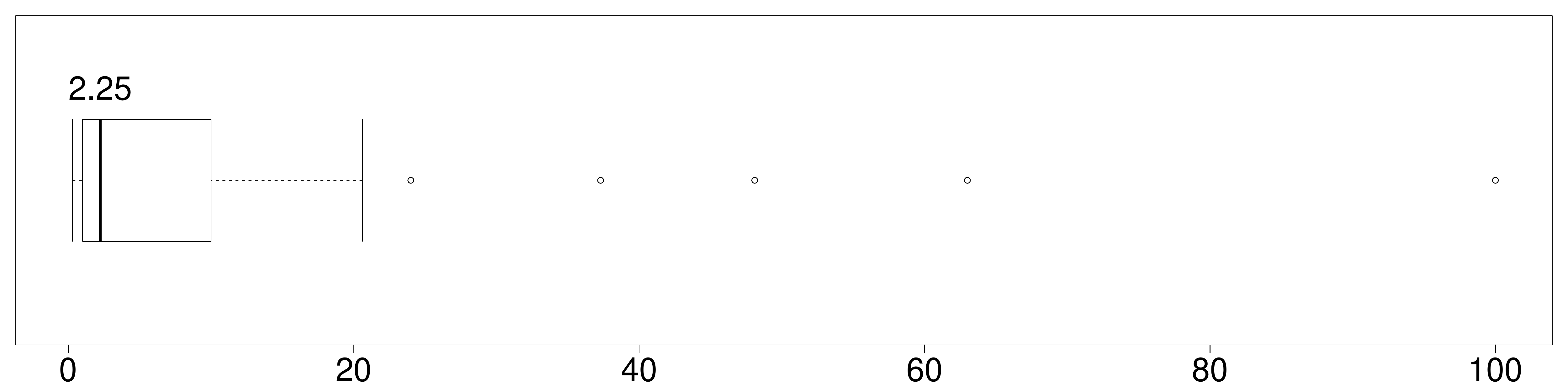}
\caption{\emph{Overlapped rate} (in \%) for each code snippet that is similar to a Stack Overflow post.}
\label{fig:overlap}
\vspace{-10pt}
\end{figure}

\noindent
\emph{\textbf{Findings.}}
We found that 232 Android code snippets are \emph{similar} to the code in Stack Overflow posts, and the Android snippets were created later than the corresponding posts. These code snippets are distributed in 135 files from 62 different apps. In other words, it is very likely that 15.5\% of the studied apps have reused code from Stack Overflow.
Figure~\ref{fig:app_clone_lines} shows the distribution of the sizes (in terms of numbers of lines of code) of the Android code snippets that were potentially reused from Stack Overflow. %are similar to code snippest And similar lines of these Android code snippets. 
The median number of lines of code in these reused code snippets is 17, which is slightly higher than NiCad's minimum line number for clones (\ie{} 10). This result suggests potential code copying from Stack Overflow to Android apps. It also shows that when developers reuse code from Stack Overflow, they take only few lines. Nevertheless, we found 25 reused code snippets with a size of more than 50 lines of code. We manually checked these large code snippets and found that:
\begin{itemize}
\item The author of the post (\#23349354) shared multiple classes on about the refreshment of the Android ListView. The \textsc{Clover} app has several methods highly \emph{similar} to the code contained in that post.
%But the app's class and the post's class are not identical.
We searched for the name of the author of this post in the app's contributor list on Github, but did not find any match. The Stack Overflow post was created in April 2014, while the corresponding Android code snippet was introduced in the app more than 10 months later (January 2015). Hence, it is very likely that the code snippet was reused from the Stack Overflow post by a developer of the app. %'s developers.
\item A developer posted a blur animation function on Android ListView (\#23844259). A \emph{similar} code snippet with more than 100 lines appeared later in the app \textsc{AcDisplay}. As in the previous case, we could not find the name of the author of the post among the contributors of the app. Also, the code was introduced in the app more than four months after it was posted on Stack Overflow. It is very likely that the developers of the app reused code from the post.
\item The \textsc{opacclient} app has a whole class \texttt{FlowLayout} highly \emph{similar} to the code contained in post \#16761418. We could not find the poster name in the list of contributors of the app. The post was created in May 2013, more than nine months before the corresponding code snippet was introduced in the app (in March 2014). Again, it is very likely that the code was reused from the post to the app.
\item Two releases of the \textsc{Clover} app contain several methods that are \emph{similar} to the code contained in post \#21857260. The post is a question about the InflateException of a custom view. It was submitted only one day before the corresponding code was introduced into the app, then the post author explained that he resolved the problem himself in the same day of the question. Therefore, we cannot confirm this case as a copy-paste operation from Stack Overflow to an app. A possible scenario could be that a developer posted some uncommitted code on Stack Overflow as a question. When the problem was resolved (by himself in this case), he applied the code in the app and committed it. %We do not match the poster's name in the app's contributor list. But the poster can use a pseudo name in Stack Overflow.
\end{itemize}

The topic of 19 out of the 25 code reuse candidates with more than 50 lines is about Android user interface. We observe that developers post problematic code on Stack Overflow when they encounter issues (such as crashes). To help other programmers understand and debug the code, the posters tend to share full classes, which also allows other developers to reuse the code to their own projects.

\hypobox{Android user interface (UI) is a hot topic on Stack Overflow. Developers reuse large UI related code snippets from Stack Overflow posts to their apps. These code snippets often contain the whole functionality of some classes.}
\vspace{5pt}

If we group the 232 code reuse candidates by their origins (\ie{} the original Stack Overflow posts), there are 45 groups where more than one candidates are related to the same Stack Overflow post. Figure~\ref{fig:apps_from_a_post} shows the distributions of the number of Android snippets (noted as $N_{Android}$) that are \emph{similar} to the same Stack Overflow post. Overall, the median number of snippets \emph{similar} to a Stack Overflow post is 1; implying that most of the Android code snippets were reused from different posts. We manually analyzed the outliers (\ie{} $N_{Android}$ > 3) in Figure~\ref{fig:apps_from_a_post}, and summarized the typical findings as follows:
\begin{itemize}
\item Several methods in the app, \textsc{WalletCordova}, were \emph{similar} to the code from post \#21907131. The \emph{similar} code is used for handling the File-transfer plugin in \textsc{PhoneGap 3.3}. The code in the Stack Overflow post and in the app is almost identical with only few modifications on variable names. It is very likely that the app's developers copied code from the post.
%\item A developer shared a whole class \texttt{FlowLayout} in the post \#16761418. A \emph{similar} class is later found in the \textsc{opacclient} app across multiple releases. We examine the contributor list of \textsc{opacclient} in Github, but do not find the poster's name. The original class accounts for 470+ lines, while the Android class accounts for 390+ lines. Though with some modifications, a number of methods in the two classes are nearly identical. This is very likely as a code reuse case.
\item The method \texttt{copyFile} contained in \textsc{Anki-Android} files is \emph{similar} to the code contained in post \#7269278. However, we did not find the author name of the post in the list of developers of the app file, or in the list of contributors to the app. The ``file copying'' method is not originally provided by Java. Thus, it is very possible that the developers of the app reused code from the post. %, and might also implement the code herself but the code happens to be similar with the code in the post.
\item The app, \textsc{frostwire-android}, possesses several methods \emph{similar} to code in post \#20027718. The poster of \#20027718 shared a whole class that can be used to customize the class \texttt{IconPageIndicator}. The app's class has multiple methods in common with the post's class, but also has some new methods that are not provided in the post. The common methods are perfectly identical. Some of the \emph{similar} methods have less than 10 lines of code, and hence were not identified by NiCad as clones, but we could identify them during our manual analysis. These similarities are a strong indication that developers may have reused code from that post to their app. %We tend to consider that the app's developers reuse code from the post.
%\item In several releases, \textsc{sophia\_oss} contains two large methods, \texttt{setListShown} and \texttt{ensureList} (60+ lines of code in total), \emph{similar} to the post \#15322624. In other methods, the app code differs from the post code.
\item \textsc{frostwire-android}, \textsc{OpenLaw}, and \textsc{reader} have a common method \texttt{setCurrentItem} which is \emph{similar} to a code snippet from the post \#14433281; implying that this method was modified and reused into different apps.
\item Several methods in the post \#8327136 have their \emph{similar} counterparts in \textsc{frostwire, OpenLaw, reader, k-9, tasks, transdroid, QuasselDroid, ad-away}, and \textsc{Atomic}. The methods were introduced in the apps later after the creation of post \#8327136. This ``popular'' post shows an example of creating a horizontal ScrollView in the Android Fragment. It is very likely that the developers of these apps borrowed code from the post.
%\item Similar to the above case, \textsc{WordPress-Android} and \textsc{Atomic} contain \emph{similar} methods to the post \#18072755.
\end{itemize}

Overall, we observe that most of the code reused candidates were reused in a single app. %Most code was ``reused'' in an app's different releases. Only a few code snippets were reused by different apps. 
Most these reused code candidates concerned general purpose issues, \eg{} how to set visual components for an Android app. Although our clone detection tool, NiCad, is set to identify only code clones that are equal or larger than 10 lines of code, we manually found some small Android code snippets (less than 10 lines of code) that are identical to the code in a post. Some long posts were reused by multiple apps or by multiple files in the same app. %In addition, most \Foutse{most? all the code snippets in the apps? what is most?} code snippets are \emph{similar} to the question post in a discussion thread. These question posts often share a large class with some minor issues in a specific method, but the whole class can work. Android developers can modify the large class for their own purpose.

%\hypobox{On the one hand, most post code was reused to a single app. On the other hand, some code was reused to an application's different releases, while a few number of code was reused by different apps. Code reused to different app files tend to derive from the question post of a discussion thread.}
%\vspace{5pt}

Figure~\ref{fig:overlap} shows the \emph{overlapped rate} (in \%) for each of the 232 code snippets that are similar to a code snippet in a Stack Overflow post. In this figure, we only depict rates that are greater than 0, which is found in 88 code snippets. The median \emph{overlapped rate} is 2.25\%. In other words, only few lines of code are both \emph{similar} to a Stack Overflow post and are contained in the license inconsistent range. %areven a single li; implying that code cloned from Stack Overflow are not strongly associated with license inconsistency.
Only for two code snippets, the \emph{overlapped rate} is greater than 50\%. However, few lines with copyright violations are enough to expose an organization to penalties. %Therefore, we do not find a strong evidence of the correlation between the code reusing candidates and the license inconsistent snippets.

%\hypobox{Although copying code from Stack Overflow is considered as one of the reasons of license incompatibility, we do not find enough evidence to support this assumption.}
%\vspace{5pt}

\subsection*{RQ2: \RQtwo}
\noindent
\emph{\textbf{Motivation.}}
Stack Overflow allows developers to reuse its content. But developers must respect the restrictions of the \emph{Creative Commons Attribute-ShareAlike 3.0 Unported} (CC BY-SA 3.0) license. Briefly, when reusing code from a Stack Overflow post, developers must cite the reference of the original post and license the resulting derivative work under CC BY-SA 3.0 or its later versions. They also need to indicate the changes if they modified the original code.
%Based on our experience, many developers do not pay enough attention to the copyright terms of a Q\&A website.
In \textbf{RQ1}, we have found 135 Android files (from 62 apps) that potentially reused code from Stack Overflow. In this research question, we intend to investigate whether the developers of these apps respected license restrictions when ``reusing'' code from Stack Overflow.

\noindent
\emph{\textbf{Approach.}}
We manually analyze the license declaration of the Android files containing code that we believe were cloned from Stack Overflow. We examine (1) whether these files use the CC BY-SA 3.0 or its later versions; (2) whether developers use CC BY-SA 3.0 or its later versions in the apps' main licenses; and (3) whether they cite the reference of the original Stack Overflow posts. If any of these conditions is not satisfied, we consider that the corresponding developers did not respect copyright terms.
To validate the results of our quantitative analysis, we send an anonymous survey to the developers of the apps that we consider as violating Stack Overflow's license. We limit the survey to apps that contain code snippets with a more than  90\% similarity (with a code snippet from Stack Overflow). We ask developers (1) whether the code snippets that we found were effectively reused from Stack Overflow, (2) whether they often reuse code from Stack Overflow, and whether they (3) consider Stack Overflow to be a reliable source of information.

\noindent
\emph{\textbf{Findings.}}
None of the 135 files that contain code potentially reused from Stack Overflow were released under CC BY-SA 3.0 or its later versions. %(the files are licensed under respectively GPL, Apache, MIT, and BSD).
And none of these files contains a reference to the corresponding Stack Overflow post (that contains the \emph{similar} code snippet). We found two posters' names in the list of contributors of the corresponding apps; indicating that the developers may have copied code from their own posts to the apps. The remaining 60 apps are therefore at risk of license violations. %s, and the organizations may have to take the corresponding legal responsibilities.
We contacted 23 developers working on these apps and received six answers. All the six developers who replied confirmed that they copied code from Stack Overflow to their projects, with one of them saying: \emph{``there is definitely code in our project that is copy-pasted from Stack Overflow, as I have done this several times. I assumed (falsely it seems) that everything there is public domain ... If I were to never look at code examples, and only write code from reading the APIs, I would probably miss elegant solutions and overlook important pitfalls.''}. Another developer said: \emph{``I often turn to StackOverflow for coding solutions ... I publish my code snippets there also ... (regarding code reuse from Stack Overflow) I would say that using code snippets `as is' usually is impossible/impractical.''} Regarding the specific code snippet that was asked, one developer replied that : \emph{``I don't remember copy-pasting code from other sites ... I actually inherited it (the cloned code) from (another project) ... I don't oppose copy-pasting to be very honest. If it was just a code snippet and I understand it, and it does what I want, I would copy-paste it to my code''}.

These results show that developers often turn to Stack Overflow for solutions. Some of them believe that reusing code from a Q\&A website like Stack Overflow can improve the quality of their software. One developer even suggested that Stack Overflow updates its license to CC BY-SA 4.0, in order to be compatible with the GPL license: \emph{``(Regarding Stack Overflow's license) it appears to not be compatible with the GNU-GPL ... I hope the staff at Stack Overflow will address the problem''}. One of the developers lamented the lack of policy about licenses in his organization: \emph{``we don't have any policy about that. Now might be a good time to have that discussion ... I have copy-pasted from Stack Overflow in the past, and still do it on projects I work on, usually with a comment citing the Stack Overflow URL''}. 

\hypobox{We recommend that software organizations put in place license control and management mechanisms, to avoid exposing themselves to license violation issues.}
\vspace{5pt}

\subsection*{RQ3: \RQthree}
\noindent
\emph{\textbf{Motivation.}}
In \textbf{RQ1} and \textbf{RQ2}, we have found evidences of code reuse from Stack Overflow to Android apps, with potential license violations. %which did not respect Stack Overflow's license.
In this question, we investigate whether developers use code from Android apps to ask or answer questions on Stack Overflow and whether they respect copyright terms when doing so. We also want to know whether there are cases of code migration to and--then from Stack Overflow, a phenomenon that we refer to as ``code laundering'' because the original license of the code would be altered by a transit on Stack Overflow. %, \ie{} an app's code was copied onto Stack Overflow then copied into another app.
%In other words, a developer shared her application's code in a Stack Overflow post. Then another developer copied the code to another application.
%If the first app did not provide a reference of the code origin, the second app can hardly apply the appropriate software license. Tracking the migration of a code snippet between Stack Overflow and Android apps can provide us a new angle of view on software license violations, because previous studies have not taken Q\&A websites into account. %The results of this question can also help developers realize the importance of copyright terms in Q\&A websites and carefully reuse code from the Internet.

\noindent
\emph{\textbf{Approach.}}
In the cloned pairs found between Stack Overflow code snippets and Android code snippets, we identify the Stack Overflow snippets that were posted later than their corresponding Android snippets' creation date. We consider that these Stack Overflow snippets reused code from an Android app. If a Stack Overflow post reused code from an app without providing the app's license, we consider it to be a license violation.
%Stack Overflow use CC BY-SA 3.0 to license its content, while most Android apps use a \emph{GPL, Apache, MIT}, or \emph{BSD} license. To the best of our knowledge, CC BY-SA 3.0 is not compatible with the above free software licenses. From the version 4.0, CC BY-SA begins to be compatible with \emph{GPL v3}~\cite{compatible_licenses}. Therefore, if developers reuse code from other people's projects under a license other than CC BY-SA 3.0 into Stack Overflow, the reusing operation could be considered as a license violation. \ANLE{Please check my arguments based on your license knowledge}
We use a Python script to automatically detect license information in the corresponding posts and manually validate our results. In addition, for each code snippet reused from an app to Stack Overflow, we examine whether the code snippet was reused later into another Android app. We also compare the software licenses of the two apps (\ie{} the source and the  destination after the transit on Stack Overflow) and check their consistency.

\begin{figure}[t]
\centering
\includegraphics[width = \columnwidth]{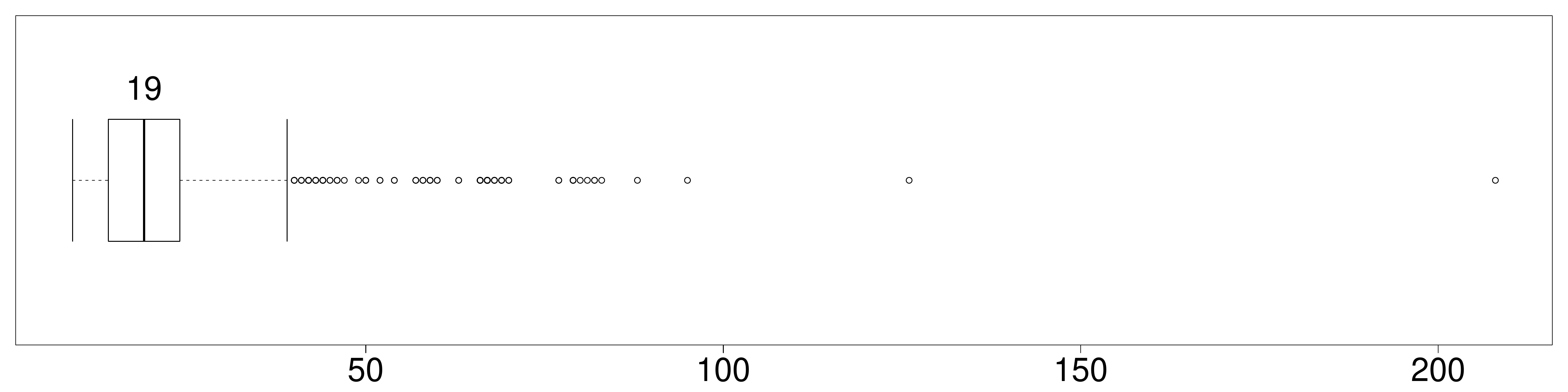}
\caption{Number of lines of code cloned from Android apps to Stack Overflow.}
\label{fig:stack_clone_lines}
\vspace{-10pt}
\end{figure}

\begin{figure}[!]
\centering
\includegraphics[width = \columnwidth]{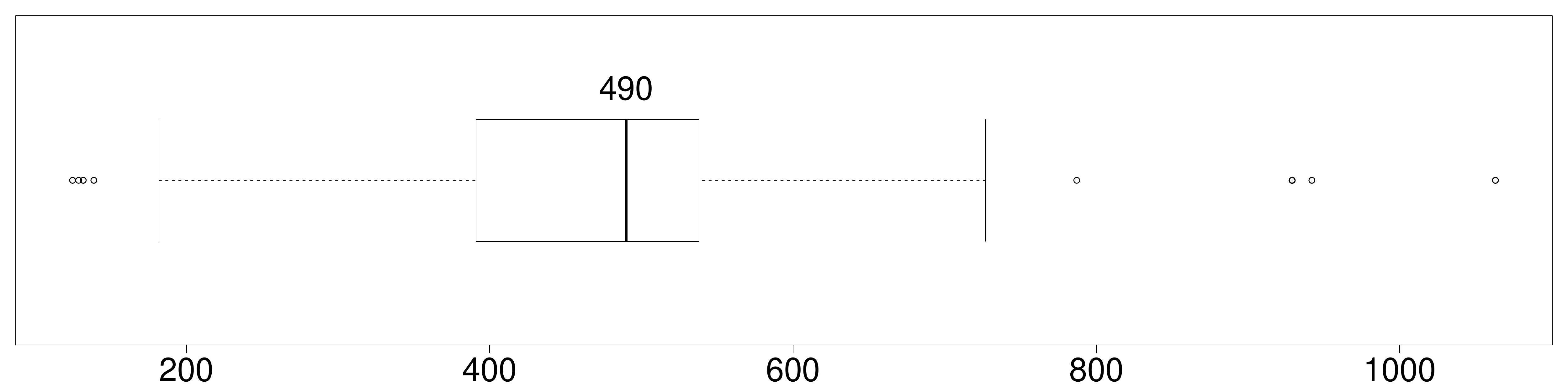}
\caption{Duration (in days) of code migration from one app to another app.}
\label{fig:migration}
\vspace{-10pt}
\end{figure}

\noindent
\emph{\textbf{Findings.}}
We found 1,226 Stack Overflow posts containing code snippets that were reused from 68 Android apps. However, only five of the posts provide the license of the original code. Although some posters claim that the code is from their own projects, we can only match two poster names in the apps' contributor lists. There is therefore a risk of license violation in the remaining 1,219 posts.
Figure \ref{fig:stack_clone_lines} shows the distribution of the sizes (in terms of numbers of lines of code) of the Stack Overflow code snippets that were potentially reused from Android apps. %distribution of line numbers of these code snippets.
The median size of these code snippets is 19 lines of code. %As aforementioned, NiCad was set to detect cloned code more than 10 lines, so the result implies that developers tend to reuse a few lines of code from Android apps to Stack Overflow.
However, we found 112 code snippets with a size of more than 50 lines of code. To understand why some developers reused such large code snippets from Android apps to Stack Overflow, we manually examined the content of the 112 posts that contain the code snippets, and made the following observations: %that contained more than 50 lines of code. These snippets are distributed in 112 posts. We read these posts and summarize the findings as follows:
\begin{itemize}
\item 107 out of the 112 code posts are related to Android UI design, with 94 posts ($94 \div 112=83.9\%$) focusing on \emph{Android Navigation Drawer}~\cite{navigation_drawer}, which is a panel that displays the app's main navigation options on the left edge of the screen~\cite{navigation_drawer}.
%In other words, some posts are about design or issues of a Navigation Drawer, others are related to other UI issues but the posters provided their Navigation Drawer code as well.
This result is surprising, because previous work (such as \cite{barua2014developers}) did not report the Navigation Drawer as a hot topic of Android. Also, in \textbf{RQ1}, we did not observe a large number of code reuses related to the Navigation Drawer from Stack Overflow to the Android apps. One explanation could be the fact that code snippets about Navigation Drawer were found only in question posts, in which it is likely that they are used as illustrations for a problem and not solutions. %that case, developers would use them to debug the answer to a question about  locate a bug debug buggy code submitted they are used  serve as Developers aims to illustrate errors in
%developers post code about Navigation Drawer when they experience a bug in that part of their code. These code examples would help developers answering their question about the errors in their code. %seek help in debugging errors
%implementing a Navigation Drawer requires a large amount of code. When encountering problems with the component, developers tend to provide entire classes to help other developers debug the code. %\ANLE{Please verify my argument}
\item Only 3 out of the 112 posts are related to general Java problems. Posts \#29242197 and \#29154598 discuss the implementation of a \texttt{java.util.Comparator}. And posts \#29242197 and \#28177863 discuss the issue of NullPointerException. %In fact, we also observe NullPointerException as a popular topic in the Android UI related posts.
\item Two posts are related to other Android problems. Post \#21299496 discusses a NullPointerException problem, while post \#34858945 discusses a file picker's problem.
\item All of the 112 posts are question posts in a discussion thread. In these questions, developers tend to share entire classes to allow other developers to test and debug their problems. However, developers did not provide any license information, in all these posts that potentially reused large code snippets from apps. %, developers do not provided the corresponding app's license. %Some of the questions seem to be posted by its corresponding app's developer, such as \#15622843. But we do not find the poster names in the apps' contributor lists. In addition, the posters did not provide the apps' licenses when they claim the code is copied from their own apps, \ie{} the posters are likely to violate the copyright terms of the apps.
These posts, which are very likely to violate the copyright terms of the apps, also impede future developers who would like to reuse the code contained in the posts, with the correct license.
\end{itemize}

\hypobox{We found 1,219 Stack Overflow posts with potential license violations. We observe that Android Navigation Drawer is a hot topic in posts that contain large code chunks reused from Android apps. We also observed that developers tend to share entire classes in the question post of a discussion thread.}
\vspace{5pt}

In our investigation of potential cases of ``code laundering'', we found 126 code snippets that first appeared in an Android app, before a code snippet exactly similar to them was posted on Stack Overflow. Later on, an exactly similar code snippet (re-)appeared in another app. We call these code snippets ``migrated code snippets''. In 12 of the migrated code snippets, the file containing the code snippet in the first app and the file containing the code snippet in the second app use different software licenses. This result shows the risk of migrating code through Stack Overflow. The license of the original code could be altered (during the migration), leading to license violations. % our assumption that migrating code snippets can lead to license incompatibility.

Figure~\ref{fig:migration} illustrates the duration (in days) of the code migrations (from one app to another app) that were found in our dataset. The median value is 490 days, \ie{} these code snippets took in average 16 months to migrate from an app to another app via Stack Overflow. The shortest migration duration is 125 days (4 months), and the longest duration is 1,063 days (35 months). %No code snippet was migrated within a very short period of time. %This result is as expected, because developers often seek a programming problem by search engines. For all of the relevant results in Stack Overflow, search engines tend to recommend the most popularly browsed posts as the top results, from which developers tend to choose their answers. However, the top posts require some time to accumulate enough click-through rate. Therefore, the migrating duration of all 126 code snippets tend to be long.

\hypobox{We found 126 code snippets that seem to have migrated from one app to Stack Overflow and then from Stack Overflow to another app. In 12 of the code snippets, the file containing the code snippet in the first app and the file containing the code snippet in the second app use different software licenses. These code snippets spent between 125 to 1,063 days to complete their migration through Stack Overflow.}
\vspace{5pt}

\subsection*{RQ4: \RQfour}
\noindent
\emph{\textbf{Motivation.}}
%In \textbf{RQ1} and \textbf{RQ2}, we find some code reusing candidates from Stack Overflow to Android apps. However, these candidates do not respect the CC BY-SA 3.0 license, \ie{} their corresponding apps have a risk of license violations.
In \textbf{RQ2} we found code snippets in Android apps that were potentially reused from Stack Overflow. The apps containing these code snippets were not released either under CC BY-SA 3.0 or its later versions. Moreover, we found no reference to the corresponding Stack Overflow posts (that contain the similar code snippets) in the apps, suggesting potential license violations. In this research question, we examine the lifespan of these code snippets in the apps to understand whether and--how developers address these potential license violation issues.

%We also observed
%In this question, we calculate the duration of a Stack Overflow code snippet remaining in released versions of an app.
%Reused code remaining in a project during a long time without appropriate license implies that developers do not realize Stack Overflow's copyright terms.

\noindent
\emph{\textbf{Approach.}}
For each app containing a code snippet that was potentially reused from Stack Overflow, we track the evolution of the code snippet across the different releases of the app, to identify the first release where the code snippet was introduced and the last release that contained the code snippet. If the code snippet was not removed from the app, the last release containing the code snippet is the latest release considered in our study. % was removed, and the code snippet still remained.
We analyze the evolution history of the app from the beginning of the project until February 2015.
 % that we believe were reused from Stack Overflow to Android apps
%For each of the 232 code reuse candidates (code snippets) from Stack Overflow to Android apps, we track its evolution in the app's releases, \ie{} we seek the first release where the code snippet was introduced, and the last release the code snippet still remained.
%For each app containing a code reuse candidate,
%
%we analyze its releases from the beginning of the project until February 2015.
Instead of considering only the code snippet's corresponding file, we take all files in the app into account. We proceed this way because a code snippet may be removed from one file and reused into another file. We use NiCad to detect duplicate code between the code snippet and each of its app's releases. In \textbf{RQ1}, we observed that multiple code snippets in the same app can be \emph{similar} to one Stack Overflow post. These code snippets are also \emph{similar} to each other. In fact, NiCad detects them as a clone class~\cite{rieger2004insights}. The 232 code snippets reused from Stack Overflow can be grouped into 124 clone classes. All the code snippets in a clone class belong to the same app. In each clone class, we identify the code snippet with the earliest creation date. We then run Nicad to detect clones between this code snippet and the files of its corresponding app's releases. %} \Foutse{you concept of earliest code snippet is not clear!!!}
%\Foutse{how many code snippets did you tracked in total?  is the following sentence accurate? "In total, we track the evolution history of 232 code snippets (found in \textbf{RQ1})"}.
%\ANLE{Please check}
Since different apps follow different release strategies (\ie{} some apps are released more frequently than other apps), we will not only report the number of releases that contain the reused code snippet, but also the duration in days from the introduction of the reused code snippet in the app until its removal or the last day of our study period. %(\ie{} February 2015).\Foutse{put the date here please!}.
%\ANLE{Please check}

\begin{figure}[t]
\centering
\includegraphics[width = \columnwidth]{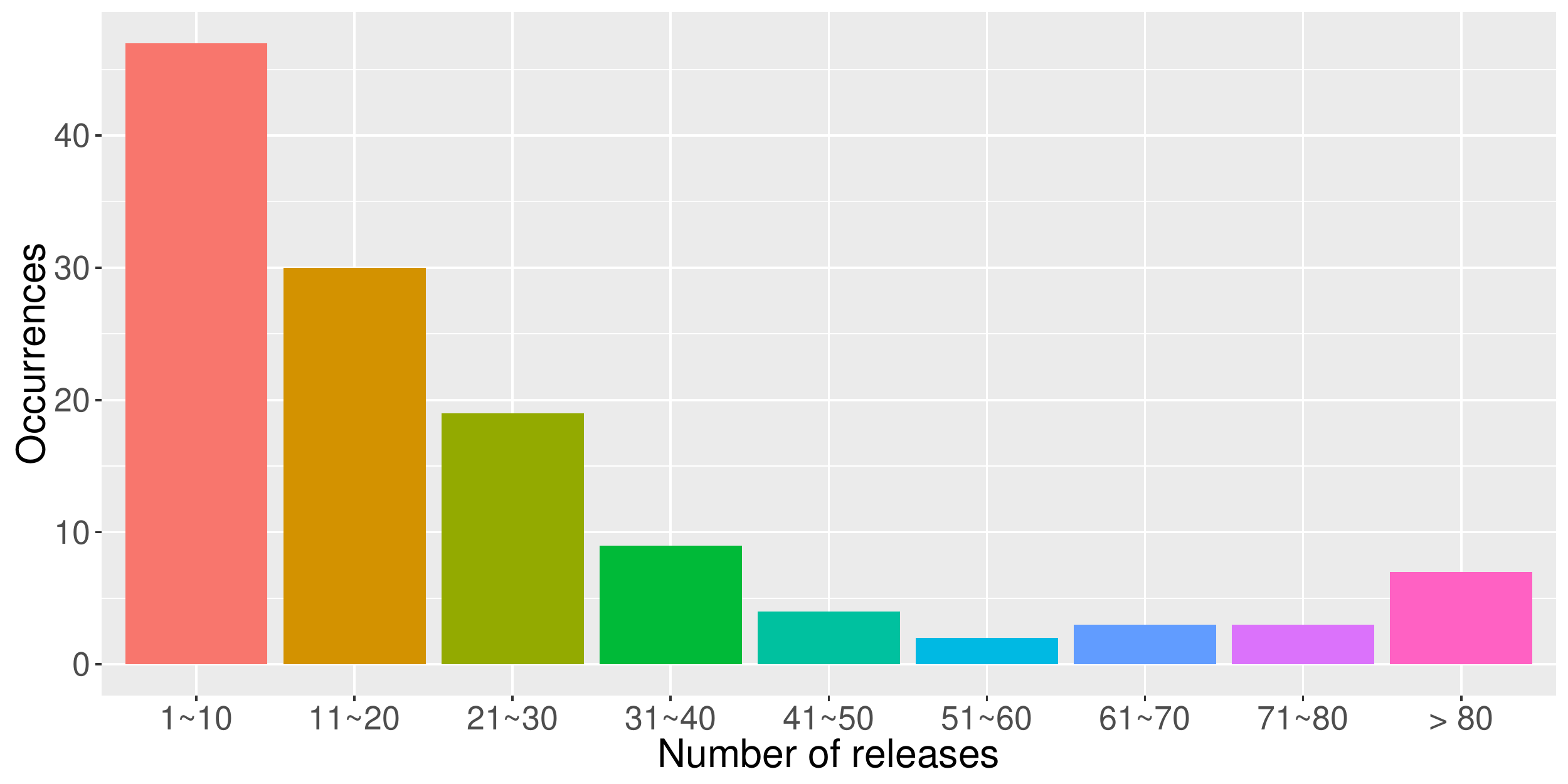}
\caption{Distribution of the numbers of released versions in which a code reuse candidate remains.}
\label{fig:remain_rel}
\vspace{-10pt}
\end{figure}

\begin{figure}[t]
\centering
\includegraphics[width = \columnwidth]{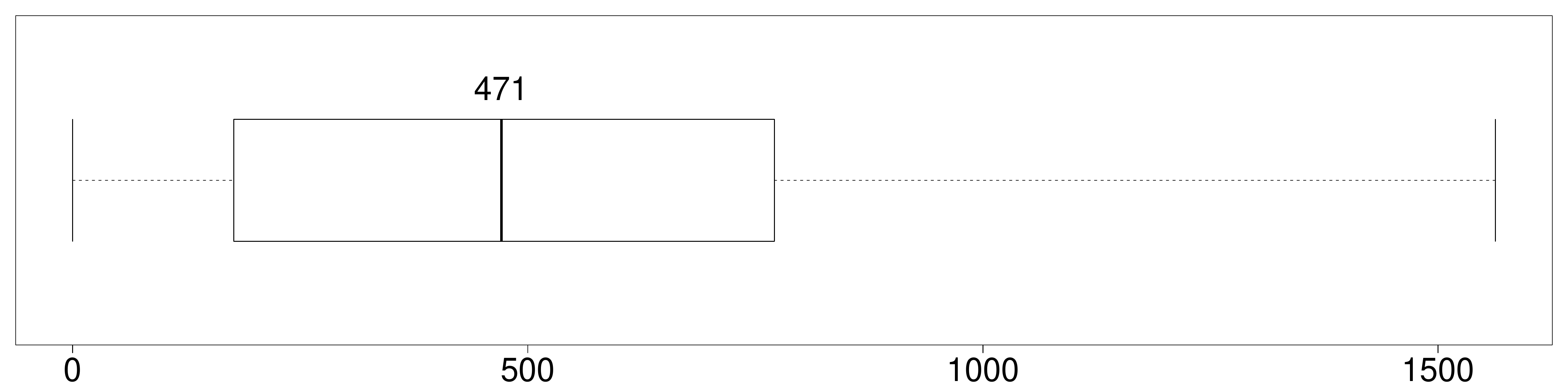}
\caption{Number of days during which a code reuse candidate remains in an app.}
\label{fig:remain_days}
\vspace{-10pt}
\end{figure}

\noindent
\emph{\textbf{Findings.}}
Figure~\ref{fig:remain_rel} shows the lifespan (in terms of number of releases) of code snippets reused from Stack Overflow to the apps.  %numbers where a code reuse candidate remains.
Figure~\ref{fig:remain_days} shows the number of days during which the code snippets reused from Stack Overflow remained in the apps. Among the 124 code snippets that were tracked through the different releases of the apps, 77 (\ie{}  62\%) remained in the app for up to 20 releases. 15 code snippets (\ie{} 12\%) remained in the app for more than 50 releases. We found five code snippets that remained in the apps for only a single release; the developers of these apps may have realized the threat posed by the copied code snippets. However, seven code snippets remained in the apps for more than 80 releases. In \texttt{Anki-Android}, we found a code snippet similar to a code snippet from Stack Overflow, that remained for 346 releases.

The median value of the lifespan (in days) of the reused code snippets is 471 days, \ie{} around 15 months. The most ephemeral code snippet stayed in its app for only 17 days, and was present in only one release (of \texttt{SafeSlinger-Android}). However, we found one code snippet similar to a code snippet from Stack Overflow in \texttt{PinDroid}. That code snippet stayed in \texttt{PinDroid} for 1,563 days (more than four years) and was present in 30 releases. %We observe that the number of releases does not always reflect a code reuse candidate's remaining days. For example, a candidate stays in \texttt{Anki-Android} for 346 releases but only during 371 days; while a candidate stays in \texttt{opentraining} for only one release but during 137 days.

Overall, the reused code snippets tend to stay in the apps for a long time, and across multiple releases. This result suggests a lack of awareness from developers, toward the risk of license violations, when reusing code from Stack Overflow. %code reuse. %  implies that developers do not realize or intentionally ignore Stack Overflow's reuse restrictions.
%for the candidates staying in multiple releases and during a long period.
%We hope that this paper can raise their awareness to the copyright terms when reusing code from the Internet, including Q\&A websites.

\hypobox{Code snippets reused from Stack Overflow tend to stay in the apps for a long time, and across multiple releases, suggesting that developers do not pay enough attention to copyright terms on Stack Overflow.}% .  Most of the code reuse candidates remain in an app within 20 releases. But some candidates can remain in an app for more than 300 releases or during more than four years. Code reuse candidates tend to remain a long time in the apps' release versions, implying developers do not realize their risk of license violations.}
\vspace{5pt}

\section{Discussion}
\label{sec:discussion}
\subsection{General Discussion}
To the best of our knowledge, this is the first quantitative study on license incompatibility between open-source software and a Q\&A website. Based on our experience, developers often reuse code from Stack Overflow in their own projects or share their projects' code to Stack Overflow. One motivation for conducting this study was the discussion that we found on Stack Exchange~\cite{license_discussion} showing developers struggling to interpret the restrictions of the CC BY-SA 3.0 license and exchanging about how to avoid license violations when reusing code from Stack Overflow. No conclusion was drawn from that discussion. A developer suggested to ``consult an attorney'' on Stack Overflow's ``specific legal issues''. In this paper, we cannot and do not intend to judge license violations from the findings of our case study. Instead, we aim to raise developers' awareness about the copyright terms on Q\&A websites.

Based on the results of our study, when reusing code from a Q\&A website, we recommend that developers provide a reference to the original code. Also, whenever it is possible, we suggest that they use a dual license (\ie{} both the license of their project and the website's license) in order to prevent license violations. When sharing code to a website, we also recommend that developers mention the license of the original project from which the code was borrowed and provide a reference to this original project. The reference can also help future developers (who reuse the code) to choose the right software license.

%\red{Before conducting this work, we interviewed some software engineers and researchers. The interviewees tend to think that copy-paste code from (or to) Stack Overflow is a common operation in development. Some of them did not know the exact copyright terms of Stack Overflow. Similarly, we also found some online discussions about the code reuse restrictions of Stack Overflow.} \ANLE{Please check whether we need to keep the above discussion.}

Although this study contains some threats to validity that we will discuss in Section~\ref{threats}, this paper sheds light on potential unethical code reuse activities taking place on Q\&A websites like Stack Overflow. % future studies in the field of software licenses, because we quantitatively find code reuse evidence from Stack Overflow to Android apps and the other way around. 
In the future, we plan to study code reuse issues in other Q\&A websites and in large-scale open-source systems. %We are also curious to know the correlation between license incompatibility and bug-proneness. %In addition, we plan to design an anonymous survey to validate the code reuse candidates detected in this study and ask the corresponding developers whether they are aware of the copyright terms of Stack Overflow.

\subsection{Threats to validity}\label{threats}
We now discuss the threats to validity of our study following the guidelines for case study research~\cite{robert2002case}.

\emph{Construct validity threats} concern the relation between theory and observation. In our study, threats to the construct validity are mainly due to measurement errors. We use the state-of-the-art clone detection tool~\cite{svajlenko2014evaluating}, NiCad, to identify similar code between the subject Android files and Stack Overflow code snippets. We use the default setting of NiCad (\ie{} minimum clone lines equal or greater than 10) to perform the clone detection, because considering code snippets with too few lines would lead to many false positives. %Also,  and too many lines can lead to false negatives in the detection. 
Nevertheless, during our manual analysis, we found some Android code snippets with less than 10 lines of code, that were highly similar or identical to a code snippet from Stack Overflow. However, the goal of this paper is not to report all similar code between the subject Stack Overflow posts and Android files. Instead, we aim to gather some evidences of code reuse activities from Stack Overflow to Android apps (and the other way around), and investigate whether developers respect copyright terms during these code reuse activities. To mitigate noises due to the precision of NiCad, for each research question, we performed manual validations. % and analysis on a selected sample.

\emph{Internal validity threats} concern factors that may affect a dependent variable and were not considered in a study. Tracking and confirming code reuse from the Internet is a very difficult task. Though we observed some Android code snippets similar or even identical to code snippets on Stack Overflow (including large code chunks), we cannot prove that the code snippets in question were ``copied'' from Stack Overflow to the apps (or the other way around). Because developers can also reuse code from other websites or open-source systems. %Therefore, the code reusing candidates identified in this paper may not be precise. 
However, the developers that we surveyed confirmed the existence of code copying from Stack Overflow to apps, \eg{} one developer stated that: \emph{``there is definitely code in our project that is copy-pasted from Stack Overflow, as I have done this several times.''}
In addition, developers often use pseudo names in their Stack Overflow accounts, which increases the difficulty of deciding whether a developer reused her own code or not. Hence, the reported license violations may actually be cases of self-copying.

%In other words, if such a candidate does not apply the appropriate license, we cannot consider it as a license violation.
%However, this paper introduces a novel direction to qualitatively study software license violations from Q\&A websites. Based on our experience, developers often seek answers from these websites, but no previous work has conducted a case study on this problem. Our findings provide potential code reuse candidates and license violation suspects, which remind developers to realize, study, and respect the copyright terms of Q\&A websites and other Internet code repositories.

\emph{Conclusion validity threats} concern the relation between the treatment and the outcome. In RQ3, we found some cases of ``code migration'' from an app ($App_A$) to Stack Overflow then to another app ($App_B$). %Because NiCad detects similar code snippets respectively in $App_A$, a Stack Overflow post, and $App_B$. 
The code snippet in $App_B$ was created later than the one in Stack Overflow, which was created later than the one in $App_A$. However, another possibility could be that both Stack Overflow and $App_B$ code snippets were reused from $App_A$. However, given the popularity of Stack Overflow, the chances that $App_B$ copied from Stack Overflow are high. We strongly recommend that developers always provide a reference and the license of the original code in their derivative works posted on Stack Overflow. This will help prevent the community from turning Stack Overflow into a ``code laundering platform''.

\emph{External validity threats} concern the possibility to generalize our results. The findings in this paper might not be generalized to other Q\&A websites and--or other systems, since our datasets were limited to some selected Android apps and Stack Overflow code snippets. Although these datasets are very large and contain apps from different domains, future studies with other open-source systems and Q\&A websites could help provide deeper insights on software license violation issues in Q\&A websites. To help researchers replicate this work or conduct future works, we share our analytic scripts and data in Github: \url{https://github.com/swatlab/stack_overflow}.

\section{Related Work}
\label{sec:related}
In this section, we discuss related works that investigated Q\&A websites and software licenses.

\subsection{Question and Answer Websites}
Q\&A websites provide a platform for users to exchange knowledge. Gy\"{o}ngyi et al.~\cite{gyongyi2007questioning} investigated user behaviours in Yahoo! Answers, which is a Q\&A website for general topics. The authors analyzed the popularity of top-level categories based on the number of questions and answers in each category.
Adamic et al.~\cite{adamic2008knowledge} investigated knowledge sharing in Yahoo! Answers. They analyzed the characteristics of the website's users and their answers, and proposed models to predict whether a particular answer will be chosen as the best answer by the asker.
%Both Gy\"{o}ngyi et al. and Adamic et al. applied \Foutse{which entropy? how?}entropy to analyze the distribution of a topic among users.

Since the introduction of Stack Overflow in 2008, a plethora of studies have focused on this Q\&A website, designed for developers.
Anderson et al.~\cite{anderson2012discovering}
%indicated \Foutse{how did they came to that conclusion? you should explain a bit their analysis....} that Stack Overflow has transitioned from a quick asking and answering website to a community-driven knowledge creation website. They
proposed models to predict the long-term value of a question and its answers on Stack Overflow. They also proposed models to predict whether a question requires a better answer.
Barua et al.~\cite{barua2014developers} explored topics and trends on Stack Overflow. They observed the growth of mobile application development questions and the decline of questions about the .NET framework. They also observed that Git has surpassed SVN in terms of the impact of version control systems,
%rise of Git as the \Foutse{what do you mean by Git becoming a popular VCS on stackoverflow....do you mean that people ask more questions about that than other VCS? please make sure the content of your sentences is clear!} most popular version control system in Stack Overflow,
and that Java is still an important programming language among developers. Vasilescu et al.~\cite{vasilescu2013stackoverflow} studied the interplay between Stack Overflow and the repository hosting website, Github. They observed that active Github committers ask fewer questions on Stack Overflow than others, and that the questions on Stack Overflow tend to be associated with the social coding in Github. However, the author did not investigate whether developers copy and paste code between the two websites and whether they respect license restrictions. Ponzanelli et al.~\cite{ponzanelli2014mining} implemented a tool, named \emph{Prompter}, which automatically retrieves pertinent answers from Stack Overflow and provides them to developers in their IDE. Through a quantitative study, they showed that \emph{Prompter} can identify the pertinent discussions if given a context in the IDE. Although this tool can provide useful suggestions, developers still need to pay attention to the licenses of the suggested code from Stack Overflow. %A corresponding license attached with each suggestion can help developers avoid license violations.
In general, none of the previous studies has investigated whether developers respect software licenses when copying code from or to Q\&A websites. Our study aims to fill this gap in the literature and raise the awareness of the software engineering community about potential unethical code reuse activities taking place on Q\&A websites. %. can alert developers and software organizations to pay attention on code reuse restrictions on Q\&A websites and to realize possible consequences due to license violations.

\subsection{Software Licenses}
Software licenses are legal instruments that govern the use or redistribution of software.
%Sojer and al.~\cite{sojer2010code} conducted a survey with 686 responses from Open Source Software (OSS) developers. They found that reusing open source code is a common phenomenon; implying that code reuse is of major importance in OSS development and has an impact on its success. they show that by performing code reuse, developers are more efficient, \eg{} developers see code reuse as a means to develop a new projects very quickly and that compete existing projects.
%However when reusing code, developers may not pay attention to the license of the code to be reused and its restrictions.

Vendome et al.~\cite{vendome2015license} conducted a study on license usage and changes in 16,221 Java projects hosted on Github. 
%, tracing \Foutse{tracing from where to where?....not sure its the right term here!} commit notes and discussions related to license changes.
They found that only 0.9\% of commit messages mention software licenses. Also, they found no discussion related to licenses in the issue reports. The authors speculated that developers are too shy to document license changes. In a follow up study~\cite{vendome2015and}, the authors indicated that developers do not necessarily know the consequences of using a specific license into their code. They conducted a survey to understand when and why developers adopt and change licenses and observed that developers have difficulties dealing with license terms (\eg{} incompatible licenses). They also observed that developers often change licenses toward more permissive licenses to facilitate the reuse of their products in commercial systems. In a recent study~\cite{mlouki2016detection}, we investigated license violations in 857 Android apps using the license detection tool Ninka~\cite{german2010sentence}, and found 399 apps with license inconsistencies (\ie{} files that share similar code but have different licenses) and 17 apps with license violations. %In this study they used Ninka a tool proposed by German and al.~\cite{german2010sentence} for license identification.}

Sojer and al.~\cite{sojer2014understanding} investigated unethical code reuse from Internet-accessible sources through a survey of 869 professional software developers. They reported that developers who perform unethical code reuse have limited knowledge on license terms and do not understand the associated legal risks. The paper suggests that software organizations warn developers about the negative consequences of unethical code reuse, provide an environment that encourages compliance with laws, and avoid excessive time pressure. %can be avoided.
Although this paper studied inappropriate code reuse from the Internet, the authors did not attempt to quantify the occurrence of unethical code reuse in real open-source projects or Q\&A websites. % nor in large-scale online code repositories. In addition, some developers may not consider using or modifying code from an answer of their questions in a Q\&A website as ``code reuse''. Hence, we address the issues in this work.

\section{Conclusion}
\label{sec:conclusion}
The question and Answer (Q\&A) website, Stack Overflow, provides a platform for programmers to share expertise and exchange ideas. It allows users to reuse its content under certain restrictions. In this paper, we examine whether developers reuse code from Stack Overflow to Android apps and whether they share code from the Android apps to Stack Overflow.
We found 232 Android code snippets that are exact clones of code snippets posted on Stack Overflow. These code snippets are distributed in 135 files from 62 different apps. We investigated the licenses of these 232 code snippets and observed potential cases of license violations in 60 apps.
%As a result, we found 62 Android apps potentially reuse code from Stack Overflow, but none of the apps respect the Stack Overflow's copyright terms.
We also found 1,226 Stack Overflow posts that potentially reused code from Android apps, and 1,219 of these posts do not respect the original apps' license. In total, we detected 1,279 cases of potential license violations.
Code snippets reused from Stack Overflow tend to stay in the apps for a long time. We also found 126 code snippets that seem to have migrated from one app to Stack Overflow and then from Stack Overflow to another app. In 12 of the code snippets, the file containing the code snippet in the first app and the file containing the code snippet in the second app used different software licenses. These findings suggest that developers do not pay enough attention to copyright terms when reusing code from Stack Overflow or sharing code on Stack Overflow. We hope that this paper will raise the awareness of the software community about potential unethical code reuse activities taking place on Q\&A websites like Stack Overflow.

%can alert developers to realize, study, and respect the copyright terms of all open-source code repositories, including Q\&A websites. In the future, we plan to study the code reusing problem in other Q\&A websites and explore the correlation between license incompatibility and bug-proneness.

%design an anonymous survey to validate the results of this paper. 

\section*{Acknowledgment}
This work is partially supported by Natural Sciences and Engineering Research Council of Canada (NSERC) and by Fonds de Recherche du Québec -- Nature et Technologies (FRQNT). We gratefully thank the developers who participated in our survey.

\balance
\bibliographystyle{IEEEtran}
\bibliography{anle_bib/anle,Libs/__reading,Libs/PURE,Libs/CASCON2009,foutsekh/foutsekh}
\end{document}